\pdfoutput=1
\documentclass[pre,aps,twocolumn,showpacs,floatfix,nofootinbib,superscriptaddress]{revtex4-1}
\usepackage{amssymb}
\usepackage{amsfonts}
\usepackage{amsmath}
\usepackage{amsthm}
\usepackage{epsfig}
\usepackage{graphicx}
\usepackage{cancel}

\usepackage{float}
\usepackage{placeins}
\usepackage{threeparttable}

\usepackage{xcolor}
\usepackage{color}
\usepackage[utf8]{inputenc}
\usepackage[hidelinks,unicode=true]{hyperref}
\usepackage{comment}
\usepackage[normalem]{ulem}
\usepackage{subcaption}
\hypersetup{colorlinks=true,% Don't draw a box around links, instead color every piece of text, which is a link
	linkcolor=blue,
	urlcolor=blue,
	citecolor=blue,
	pdfhighlight=/N% When clicking a link and holding the mouse button, don't use any graphical effects - i.e., the alternative would be to behave like a "button" which is pressed within the PDF
}
	\renewcommand{\vec}{\mathbf}

	\usepackage{ragged2e} % para \justifying
	
	\makeatletter
	\long\def\@makecaption#1#2{%
		\vskip\abovecaptionskip
		\begingroup
		\justifying              % <<--- texto justificado
		\small                   % tamanho da legenda (mude se quiser)
		\parindent=0pt
		\textbf{#1.}~#2\par      % "FIG. 3." em negrito + texto
		\endgroup
		\vskip\belowcaptionskip
	}
	\makeatother

\begin{document}
%%%%%%%%%%%%%%%%%%%%%%%%%%%%%%%%%%%%%%%%%%%%%%%%%%%%%%%%%%%%%%%%%%%
%\title{ \textcolor{red}{Scaling, Fractal Dynamics and Critical Exponents in Non-Integer Dimensional Ising Model}}
%\title{ Scaling, Fractal Dynamics and Critical Exponents: Application in a non-integer dimensional ising model}
%\title{ Phase Transition in disordered system: Fractal Dynamics, Scaling and Critical Exponents}
\title{Fractal signatures of strong universality in the disordered Ising model}
\author{Henrique A Lima}
\email{Henrique_adl@hotmail.com}
\address{International Center of Physics, Institute of Physics, University of Brasilia, 70910-900, Brasilia, Federal District, Brazil}
\author{Kaue Hermann}
\email{kaueabbehausen@hotmail.com}
\address{International Center of Physics, Institute of Physics, University of Brasilia, 70910-900, Brasilia, Federal District, Brazil}
\author{Ismael S. S. Carrasco}
\email{ismael.carrasco@ufv.br}
\address{Departamento de F\'isica, Universidade Federal de Vi\c cosa, 36570-900, Vi\c cosa, MG, Brazil}
\author{Jairo R. L. de Almeida}
\email{jairorolimalmeida19@gmail.com}
\address{Instituto de F\'{\i}sica, Universidade Federal de Pernambuco,  Recife, PE, Brazil}
%\author{Marcelo Lyra}
%\email{marcelo@fis.ufal.br }
%\address{Instituto de F\'{\i}sica, Universidade Federal de Alagoas,  Macei\'{o}, AL, Brazil}
\author{Fernando A. Oliveira}
\email{faooliveira@gmail.com}
\address{International Center of Physics, Institute of Physics, University of Brasilia, 70910-900, Brasilia, Federal District, Brazil}
%\address{Instituto de F\'{\i}sica, Universidade Federal Fluminense, Avenida Litor\^{a}nea s/n, 24210-340, Niter\'{o}i, RJ, Brazil}
%%%%%%%%%%%%%%%%%%%%%%%%%%%%%%%%%%%%%%%%%%%%%%%%%%%%%%%%%%%%%%%%%%%
\begin{abstract}
  Disordered magnetic systems provide a valuable setting for investigating how quenched disorder affects critical behavior. In this work, we study a two-dimensional random-bond Ising model with Gaussian-distributed couplings of positive mean and analyze how the estimates of thermodynamic and geometrical quantities depend on the disorder strength. Since bond disorder is marginal in the two-dimensional Ising universality class, individual critical exponents obtained from finite numerical data may be strongly affected by logarithmic corrections and by the choice of fitting windows. We therefore compare the disorder dependence of the usual exponent estimates with that of exponent ratios and fractal dimensions. In this context, the correlation function of order-parameter fluctuations, introduced by Fisher, provides a central framework for describing second-order phase transitions at equilibrium. At criticality, the anomalous exponent $\eta$ corrects the Euclidean scaling of the correlation function. Recent works showed that this correction can be interpreted geometrically through a fractal description of critical spin clusters, leading to relations between $\eta$, the Riesz fractal dimension $d_R$, and the fractal dimension $d_f$ of the ordered phase. Here, we use Monte Carlo simulations to test the robustness of these geometrical quantities in the presence of quenched bond disorder. Our results show that the direct estimates of exponents such as $\beta$, $\nu$, and $\gamma$ display noticeable disorder-dependent variations, consistent with the expected difficulty of extracting asymptotic exponents in marginally disordered systems. In contrast, the ratio $\beta/\nu$ as well as the fractal dimensions $d_f$ and $d_R$, remain much more stable and close to their pure two-dimensional Ising values. These results support the strong-universality scenario with logarithmic corrections and suggest that fractal dimensions provide a robust geometrical route for identifying the underlying universality class in disordered critical systems. 
\end{abstract}
%%%%%%%%%%%%%%%%%%%%%%%%%%%%%%%%%%%%%%%%%%%%%%%%%%%%%%%%%%%%%%%%%%%
\maketitle
%%%%%%%%%%%%%%%%%%%%%%%%%%%%%%%%%%%%%%%%%%%%%%%%%%%%%%%%%%%%%%%%%%%
\section{Introduction}
\label{Int}

Disordered magnetic systems provide a fundamental setting for investigating how quenched randomness affects collective behavior and critical phenomena. In lattice spin systems, disorder may appear in different forms, such as random fields, site dilution, bond dilution, or spatial fluctuations in the exchange couplings \cite{Elliott74,Fytas2013}. These different types of disorder may lead to distinct physical regimes, ranging from diluted ferromagnets to frustrated spin glasses and percolation transitions \cite{Elliott74,Binder86,Cardy84}. In the present work, we focus on a ferromagnetic random-bond Ising model in two dimensions, where disorder is introduced through spatial fluctuations of the nearest-neighbor couplings.

A general Hamiltonian for an Ising system with quenched magnetic disorder can be written as
\begin{equation}
	\label{Ham}
	H(s) = -\sum_{\langle i,j\rangle}J_{i,j}s_{i}s_{j}
	-\sum_{i}h_i s_{i},
\end{equation}
where $s_i=\pm 1$ is the spin variable at site $i$, $J_{i,j}$ is the coupling between nearest-neighbor spins, and $h_i$ is a local magnetic field. Randomness in $h_i$ leads to random-field models \cite{Fytas2013}, whereas randomness in $J_{i,j}$ leads to random-bond models \cite{Fytas2010a}. When the coupling distribution contains a significant fraction of positive and negative bonds, frustration may become relevant and the system may approach the spin-glass regime described by the Edwards-Anderson model \cite{Edwards75,Binder86}. Here, however, we consider the case $h_i=0$ and Gaussian-distributed couplings with positive mean,
\begin{equation}
	\label{dis}
	P(J_{i,j})=
	\frac{1}{\sqrt{2 \pi\sigma^2 }}
	\exp\left[-\frac{(J_{i,j}-J)^2}{2 \sigma^{2}}\right].
\end{equation}
The average coupling is $J=\langle J_{i,j}\rangle$, and $\sigma$ controls the disorder strength. In this work we take $J>0$, so that ferromagnetic order is favored. Therefore, for the range of disorder investigated here, the system remains in the ferromagnetic random-bond regime and should not be interpreted as a spin-glass system. The couplings are quenched and symmetric, $J_{i,j}=J_{j,i}$, and the standard two-dimensional Ising model is recovered in the limit $\sigma=0$.

The influence of quenched bond disorder on the critical behavior of the two-dimensional Ising model has been extensively investigated over the past few decades. According to the Harris criterion, disorder is relevant when the specific-heat exponent of the pure system is positive, irrelevant when it is negative, and marginal when it vanishes \cite{Harris74}. Since the pure two-dimensional Ising model has $\alpha=0$ \cite{Cardy96}, bond disorder represents a marginal perturbation. This marginality led to a long-standing debate between two scenarios. In the weak-universality scenario, individual exponents such as $\beta$, $\nu$, and $\gamma$ may vary with disorder \cite{Suzuki74,Fytas08}. In the strong-universality scenario, the leading critical exponents remain those of the pure Ising model, whereas disorder produces logarithmic corrections to the asymptotic scaling laws. Distinguishing between these scenarios numerically has typically required advanced methods or detailed scaling analyses, including transfer-matrix calculations, Wang--Landau sampling, and finite-size scaling \cite{Reis96,Reis97,Fytas08,Fytas10}. These studies generally favor the strong-universality scenario with logarithmic corrections. This motivates the search for observables that can identify the asymptotic universality class more directly, without requiring advanced methods to estimate individual critical exponents.

This issue is particularly important from a numerical point of view. If the asymptotic behavior contains multiplicative logarithmic corrections, quantities measured over finite temperature intervals or finite system sizes may be well described only by effective exponents \cite{Reis96,Reis97,Fytas10}. For example, a singular quantity $X$ may behave as
\begin{equation}
	X(t)\sim t^{-x}|\ln t|^{\hat{x}},
\end{equation}	%CHECAR ESTA EQUAÇÃO
where $t=|T-T_c|/T_c$ is the reduced temperature. If the logarithmic factor is not explicitly included, fitting the data with a pure power law may produce disorder-dependent estimates of $x$, even when the leading exponent remains unchanged. Thus, apparent variations of $\beta$, $\nu$, or $\gamma$ with $\sigma$ should be interpreted with caution. In this context, exponent ratios may provide more stable numerical indicators of the underlying universality class \cite{Reis97,Fytas08}, motivating the comparison with geometrical observables pursued here.

Phase transitions are characterized by singular changes in thermodynamic and correlation functions as external parameters, such as temperature or pressure, are varied \cite{Kardar07}. For the Ising model, the order parameter is the magnetization
\begin{equation}
	\label{m}
	m = \frac{1}{N}\sum_{k}s_{k},
\end{equation}
where the sum runs over all $N$ spins. The fluctuation of the order parameter, $\psi(\vec{x})=m-s(\vec{x})$, contains important information about correlations near criticality. In particular, the correlation function
\begin{equation}
	\label{G0}
	G(r)=\langle \psi(\vec{r}+\vec{x})\psi(\vec{x})\rangle
\end{equation}
is a central quantity in the theory of second-order phase transitions. Here, $\langle \cdots\rangle$ denotes an average over the lattice. In the long-wavelength limit, the Ornstein-Zernike description leads to \cite{Ornstein14,Kardar07}
\begin{equation}
	\label{G}
	(-\nabla^2 +\rho^{-2})G(r)=\delta^{(d)}(r),
\end{equation} 
where $\rho$ is the correlation length. Close to the critical point, the correlation length diverges as \cite{Kardar07,Cardy96}
\begin{equation}
	\label{rhodivergence}
	\rho \propto |T-T_c|^{-\nu}.
\end{equation}
The solution of Eq. (\ref{G}) gives $G(r)\propto r^{2-d}\exp(-r/\rho)$. At criticality, where $\rho\rightarrow\infty$, this expression predicts $G(r)\propto r^{2-d}$.

However, this classical result does not correctly describe the critical correlations of the Ising model. In two dimensions, it would predict a distance-independent correlation function, whereas the exact result of Kaufman and Onsager gives $G(r)\propto r^{-1/4}$ at criticality \cite{Kaufman49}. More generally, Fisher introduced the anomalous exponent $\eta$ through~\cite{Fisher64}
\begin{equation}
	\label{G2}
	G(r) \propto r^{2-d-\eta}.
\end{equation}
The exponent $\eta$ therefore encodes the anomalous spatial scaling of critical fluctuations.

At a second-order phase transition, scale invariance also implies the emergence of fractal structures \cite{Suzuki83,Coniglio89}. In magnetic systems, the critical spin configurations are organized into clusters whose geometry is related to the usual thermodynamic exponents. For instance, the fractal dimension $d_f$ of the ordered phase satisfies the scaling relation \cite{Suzuki83}
\begin{equation}
	\label{df}
	d_f=d-\frac{\beta}{\nu}.
\end{equation}
For percolation, the corresponding fractal object is the infinite percolating cluster \cite{Grimmett06,Cruz23}, while in magnetic systems the relevant critical clusters are related to the thermodynamic behavior \cite{Suzuki83,Coniglio89}. Equation (\ref{df}) shows that the ratio $\beta/\nu$, rather than the individual values of $\beta$ and $\nu$, directly controls this geometrical exponent. This is relevant for disordered systems, where individual exponent estimates may be strongly affected by logarithmic corrections, while exponent ratios can be numerically more stable \cite{Reis97,Fytas08}.

Recently, a geometrical interpretation of the Fisher exponent was proposed using fractional calculus and fractal derivatives \cite{Lima24,Lima25,Carrasco26}. The central idea for correcting the fluctuation-dissipation theorem (FDT) violation\footnote[1]{Note that, violation in the susceptibility behavior is a violation in the   FDT~\cite{GomesFilho25}} is that, in criticality, the relevant variations of the correlation function occur in a scale-invariant fractal structure, and not in the entire d-dimensional Euclidean space. In this case, the Euclidean operator in Eq. (\ref{G}) may be replaced by a fractional Riesz operator,
\begin{equation}
	\label{G3}
	(-\nabla^2)^\zeta G(r)=\delta^{(d_R)}(r),
\end{equation}
where $d_R$ is a fractal dimension associated with the Riesz derivative of order $\zeta$. This approach leads to the relation
\begin{equation}
	\label{eta}
	\eta=d-d_R=1-\zeta.
\end{equation}
Furthermore, Lima {\it et al.} \cite{Lima24} obtained the relation
\begin{equation}
	\label{dR}
	d_R=2(d_f-1).
\end{equation}
Together, Eqs. (\ref{df}), (\ref{eta}), and (\ref{dR}) connect the anomalous decay of correlations with the geometry of critical clusters.

The main objective of this work is to investigate how thermodynamic, correlation, and geometrical quantities behave in the two-dimensional random-bond Ising ferromagnet as the disorder strength $\sigma$ is varied. Our central result is that, while the direct estimates of $\beta$, $\nu$, and $\gamma$ show noticeable variations with disorder, the ratios $\beta/\nu$ and $\gamma/\nu$ and the fractal dimensions $d_f$ and $d_R$ remain much more stable and close to the exact values of the pure two-dimensional Ising model. In particular, the direct estimate of $d_R$ shows no systematic drift with disorder and remains compatible with $d_R=7/4$. This supports the strong-universality scenario and suggests that $d_R$ provides a comparatively simple and  robust route for identifying the underlying universality class in marginally disordered critical systems.

%%%%%%%%%%%%%%%%%%%%%%%%%%%%%%%%%%%%%%%%%%%%%%%%%%%%%%%%%%%%%%%%%%%

\section{Simulations in a disordered lattice}

Fisher also proposed the scaling relation~\cite{Fisher64}
\begin{equation}
	\label{gamma}
	\gamma =(2-\eta)\nu,
\end{equation}
where $\gamma$ is the susceptibility exponent. Another important relation connects the correlation-length exponent $\nu$ to the specific-heat exponent $\alpha$ through the hyperscaling relation~\cite{Cardy96}
\begin{equation}
	\label{alpha}
	\alpha=2-d\nu.
\end{equation}
Together with the Rushbrooke equality~\cite{Cardy96},
\begin{equation}
	\label{rush}
	\alpha +2\beta +\gamma=2,
\end{equation}
these relations provide useful consistency checks for the critical behavior. 

Before proceeding with the numerical analysis, we recall that fractal exponents also appear in the description of percolation clusters, as discussed, for instance, by Suzuki~\cite{Suzuki83} and Coniglio~\cite{Coniglio89}. The exponent $d_R$ introduced here represents a different fractal dimension, associated with the Riesz description of the correlation function \cite{Lima24}. Nevertheless, Eq. (\ref{dR}) establishes a direct connection between $d_R$ and the fractal dimension $d_f$ of the ordered phase.

We now turn to the numerical analysis of the disordered lattice. We consider the Hamiltonian in Eq. (\ref{Ham}) with $h_i=0$ on a square lattice of lateral size $L$, with $N=L^2$ spins. The couplings are drawn from the Gaussian distribution in Eq. (\ref{dis}), with fixed mean value $J=1$, while the standard deviation $\sigma$ controls the disorder strength. Since the distribution has a positive mean and the values of $\sigma$ used here satisfy $\sigma/J\leq 0.5$, the simulated systems remain in the ferromagnetic random-bond regime.

For $\sigma=0$, the model reduces to the pure two-dimensional Ising model, whose exact critical exponents are $\beta=1/8$, $\nu=1$, and $\eta=1/4$~\cite{Cardy96,Kaufman49}. The corresponding geometrical values are $d_R=7/4$ for the Riesz fractal dimension~\cite{Lima24,Lima25} and $d_f=15/8$ for the critical-cluster fractal dimension\cite{Suzuki83,Coniglio89}. The central question addressed below is how stable these quantities remain when quenched bond disorder is introduced.

\begin{figure}[h!]
	\centering
	\includegraphics[width=1\linewidth]{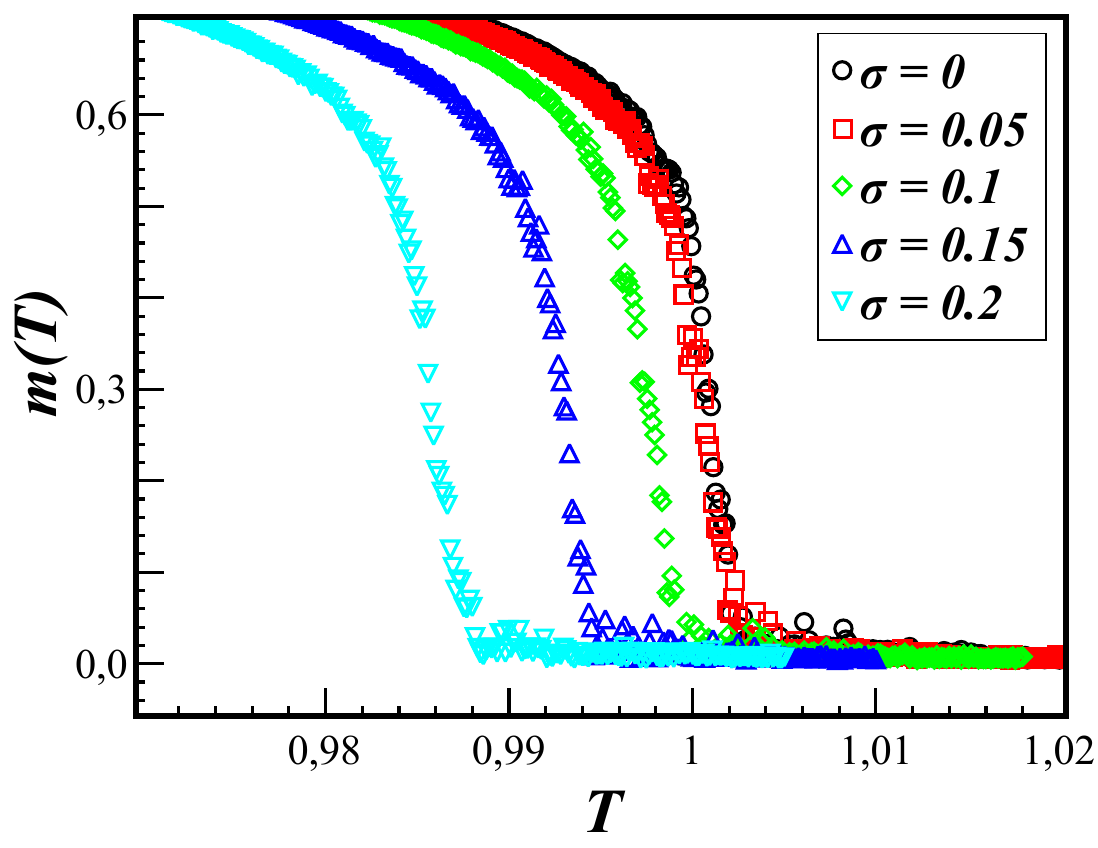}
	\caption{Order parameter $m$ as a function of the normalized temperature $T/T_O$ for different values of the disorder strength $\sigma$.}
	\label{fig:mag1}
\end{figure}

\begin{figure}[h!]
	\centering
	\includegraphics[width=1\linewidth]{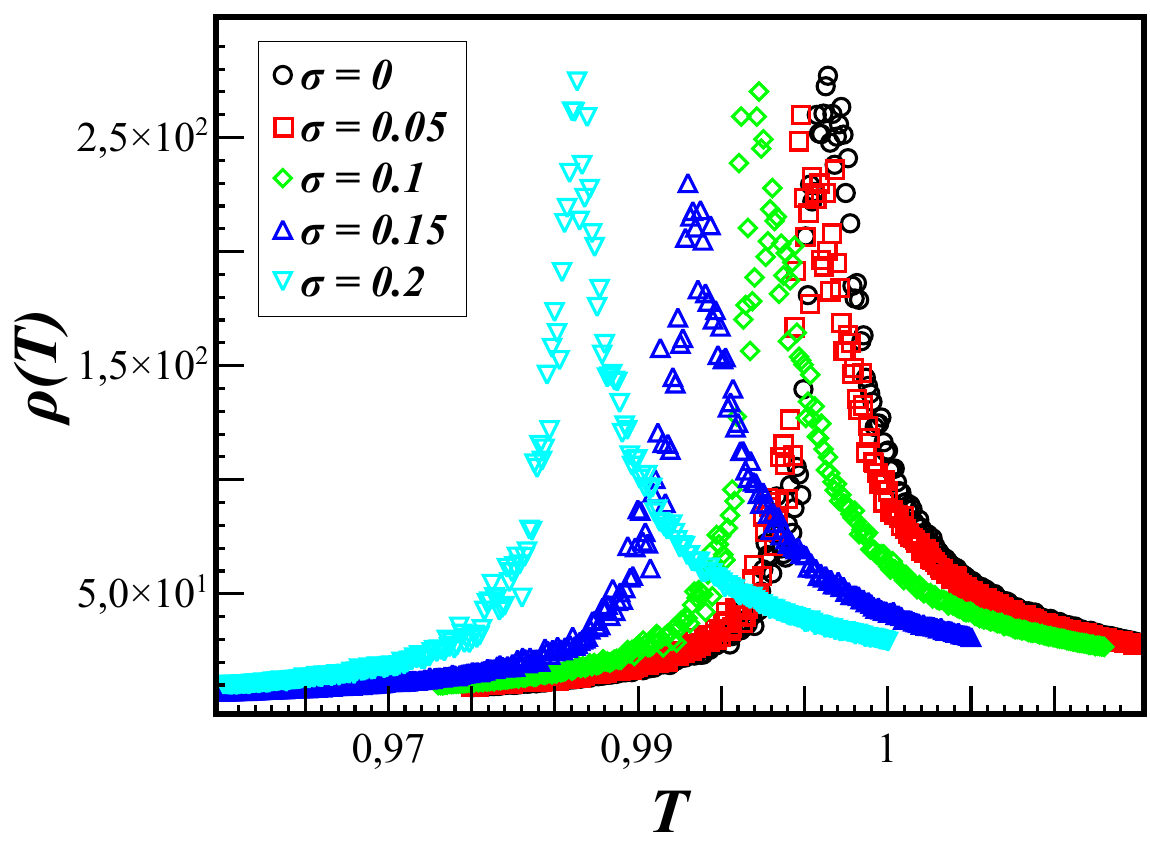}
	\caption{Correlation length $\rho$ as a function of the normalized temperature $T/T_O$ for different values of the disorder strength $\sigma$.}
	\label{fig:rho}
\end{figure}

\begin{figure}[h!]
	\centering
	\includegraphics[width=0.98\linewidth]{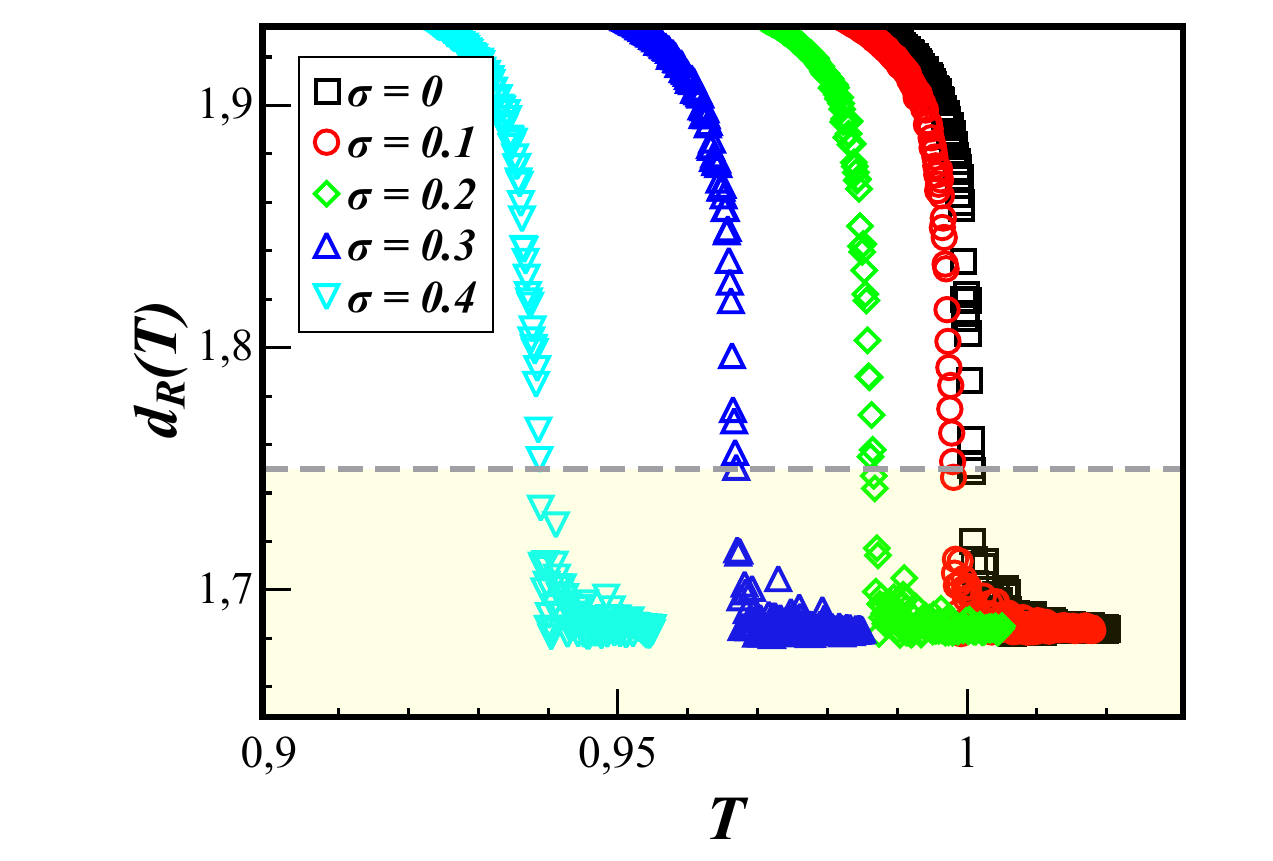}
	\caption{Riesz fractal dimension $d_R$ as a function of the normalized temperature $T/T_O$ for different values of the disorder strength $\sigma$. The dashed line indicates the pure Ising value $d_R=7/4$.}
	\label{fig:dr}
\end{figure}

The simulations shown in Figs. \ref{fig:mag1}--\ref{fig:dr} were performed on lattices with lateral size $L=4096$. To simulate systems of this size, we used \textit{CUDA} technology for parallel computation. The Ising model was evolved using a Monte Carlo algorithm with a checkerboard update scheme, in which the lattice is divided into two sublattices. Spins on one sublattice are updated while the other sublattice is kept fixed, and one Monte Carlo sweep corresponds to the sequential update of both sublattices. For each temperature $T$ we performed $3 \times 10^{5}$ Monte Carlo sweeps. This procedure is well suited for GPU parallelization and allows the analysis of large lattice configurations, which is particularly useful for extracting geometrical quantities by box-counting methods \cite{Binder12,Romero_2020}.

In Figs. \ref{fig:mag1}--\ref{fig:dr}, we show the temperature dependence of the magnetization $m$, the correlation length $\rho$, and the Riesz fractal dimension $d_R$ for different values of $\sigma$. The temperature is measured in units of the Onsager critical temperature $T_O = 2.269J/k_B$, which is the critical temperature of the pure model. As shown in Fig. \ref{fig:mag1}, the transition temperature decreases as the disorder strength increases. To estimate the transition temperature and the order-parameter exponent, the magnetization data were fitted near the transition using
\begin{equation}
	\label{box1}
	m(T) =
	\begin{cases}
		a|T_{c}-T|^\beta, &\text{ if } T < T_c, \\
		0, &\text{ if } T \geq T_c.
	\end{cases}
\end{equation}
In a marginally disordered system, this expression should be understood as the leading power-law part of a more general scaling form that may include multiplicative logarithmic corrections \cite{Reis96,Reis97}, which are not included here, since our objective is to examine the behavior of the fractal dimensions using a simple analysis. Consequently, the exponent $\beta$ extracted from Eq. (\ref{box1}) is an effective exponent, dependent on the accessible temperature range and on the fitting procedure. The critical temperature obtained from this analysis is denoted by $T_c^{\mathrm{mag}}$.

Figure \ref{fig:rho} shows the correlation length $\rho(T)$ as a function of temperature for different values of $\sigma$. Close to the transition, the leading critical behavior is expected to follow Eq. \ref{rhodivergence}. As in the case of the magnetization, the values of $\nu$ obtained from pure power-law fits should be interpreted as effective estimates. In a marginally disordered system, logarithmic corrections and finite-size effects make the extrapolation to the thermodynamic critical regime particularly delicate \cite{Reis96,Reis97,Fytas10}. Therefore, fits performed at a fixed lattice size and over finite temperature intervals may produce apparent disorder-dependent exponents even if the asymptotic exponent remains unchanged.

Figure \ref{fig:dr} shows the Riesz fractal dimension $d_R(T)$ as a function of temperature. The values of $d_R$ were obtained using a parallel box-counting method, in which the relevant critical set is covered by boxes of different sizes and the scaling of the number of occupied boxes is analyzed~\cite{Lima24}. 

The values of $T_c^{\mathrm{mag}}$, $T_c^{d_R}$, and the effective exponents are summarized in Table \ref{Table1}. The small differences between $T_c^{\mathrm{mag}}$ and $T_c^{d_R}$ provide an estimate of the systematic uncertainty associated with the two procedures. These differences become more visible as $\sigma$ increases, which is consistent with the expected difficulty of locating the transition in the presence of marginal disorder and logarithmic corrections \cite{Reis96,Reis97}.

A key point in Table \ref{Table1} is that the individual exponent estimates display noticeable variations with $\sigma$. This should not be interpreted as direct evidence for continuously varying critical exponents or for weak universality. Instead, these variations are naturally understood as effective-exponent effects caused by finite fitting windows and by the logarithmic corrections expected for the two-dimensional random-bond Ising model \cite{Reis96,Reis97,Fytas10}. In contrast, the ratio $\beta/\nu$ remains close to the pure Ising value $1/8$ over the whole range of disorder investigated. This stability supports the strong-universality scenario and motivates the use of geometrical quantities, such as $d_f$ and $d_R$, as they are functions of this ratio.

\begin{table*}[htbp]
\centering
\begin{tabular}{cccccccccc}
\hline\hline
$\sigma$ & $T_c^{\mathrm{mag}}$ & $T_c^{d_R}$ &
$\beta$ & $\nu$ & $\beta/\nu$ & $\gamma$ & $\gamma/\nu$ &
$\alpha$ & $\Sigma$ \\
\hline
0.00 & 1.0000(2)    & 1.0007(5)   & 0.1253(6)  & 1.000(5)  & 0.1252(8)  & 1.749(10) & 1.7496(16) & -0.0016(99) & 1.998(14) \\
0.05 & 0.99901(8)   & 1.00035(12) & 0.124(1)   & 0.992(7)  & 0.1250(13) & 1.736(14) & 1.7500(26) & 0.016(14)   & 2.000(20) \\
0.10 & 0.99607(3)   & 0.99683(14) & 0.121(1)   & 0.971(8)  & 0.1246(15) & 1.700(16) & 1.7508(30) & 0.058(16)   & 2.000(23) \\
0.15 & 0.9912(2)    & 0.9925(4)   & 0.1211(7)  & 0.976(12) & 0.1241(17) & 1.710(24) & 1.7518(34) & 0.048(24)   & 2.000(34) \\
0.20 & 0.98448(3)   & 0.9860(8)   & 0.1216(9)  & 0.981(15) & 0.1240(21) & 1.719(30) & 1.7520(42) & 0.04(3)     & 2.002(42) \\
0.25 & 0.9754(7)    & 0.97704(23) & 0.1229(2)  & 0.977(6)  & 0.1258(8)  & 1.708(12) & 1.7484(16) & 0.046(12)   & 2.000(17) \\
0.30 & 0.96419(3)   & 0.9655(8)   & 0.121(1)   & 0.963(12) & 0.1256(19) & 1.684(24) & 1.7488(38) & 0.074(24)   & 2.000(34) \\
0.35 & 0.95088(4)   & 0.95308(20) & 0.1209(15) & 0.962(11) & 0.1257(21) & 1.682(22) & 1.7486(42) & 0.076(22)   & 2.000(31) \\
0.40 & 0.934944(11) & 0.9380(6)   & 0.119(4)   & 0.958(24) & 0.124(5)   & 1.678(49) & 1.752(10)  & 0.08(5)     & 1.996(69) \\
0.45 & 0.9162(3)    & 0.9201(4)   & 0.119(2)   & 0.968(14) & 0.123(3)   & 1.698(28) & 1.754(6)   & 0.064(28)   & 2.000(40) \\
0.50 & 0.8962(2)    & 0.90025(13) & 0.122(2)   & 0.967(30) & 0.126(4)   & 1.690(60) & 1.748(8)   & 0.07(6)     & 2.004(85) \\
\hline\hline
\end{tabular}

\caption{Critical temperatures $T_c^{\mathrm{mag}}$ and $T_c^{d_R}$, effective exponent estimates, and exponent ratios as functions of the disorder strength $\sigma$. Here, $T_c^{\mathrm{mag}}$ is obtained together with $\beta$ from the critical fit of the magnetization, $m(T)\propto (T_c^{\mathrm{mag}}-T)^\beta$, whereas $T_c^{d_R}$ is obtained from a local polynomial fit, Eq.~(\ref{drT}), of $d_R(T)$ around $d_R=7/4$ and is defined as the temperature at which the fitted curve satisfies $d_R=7/4$. The exponents $\beta$ and $\nu$ are obtained from direct power-law fits, while $\alpha$ and $\gamma$ are obtained from Eqs.~(\ref{alpha}) and (\ref{gamma}), respectively. The values should be interpreted as effective estimates because logarithmic corrections are expected in the two-dimensional random-bond Ising model. The approximate invariance of $\beta/\nu$ and $\gamma/\nu$ is consistent with the pure Ising values $1/8$ and $7/4$, respectively, and supports the strong-universality scenario. Here, $\Sigma=\alpha+2\beta+\gamma$ tests the Rushbrooke relation in Eq.~(\ref{rush}).}
\label{Table1}
\end{table*}

\subsection{Data analysis}

\begin{figure}[!]
	\centering
	% Painel (a)
	\begin{subfigure}[c]{0.83 \linewidth}
		\centering
		\includegraphics[width=\linewidth]{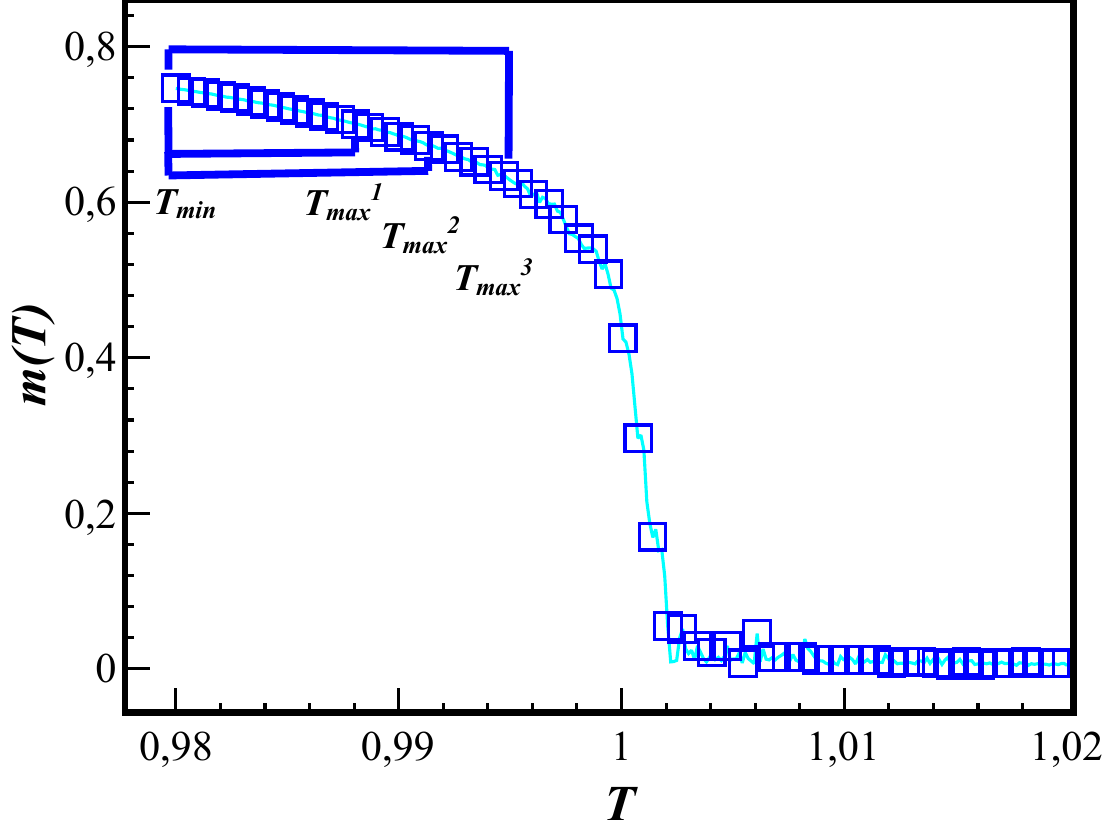}
		\caption{}
		\label{mag_tc_tmax}
	\end{subfigure}
	% Painel (b)
	\begin{subfigure}[c]{0.88 \linewidth}
		\centering
		\includegraphics[width=\linewidth]{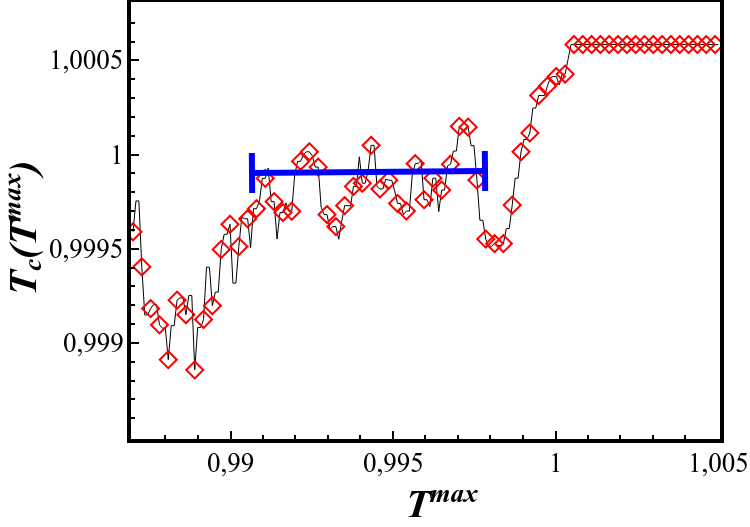}
		\caption{}
		\label{tc_tmax}
	\end{subfigure}
	% Painel (c)
	\begin{subfigure}[c]{0.92 \linewidth}
		\centering
		\includegraphics[width=\linewidth]{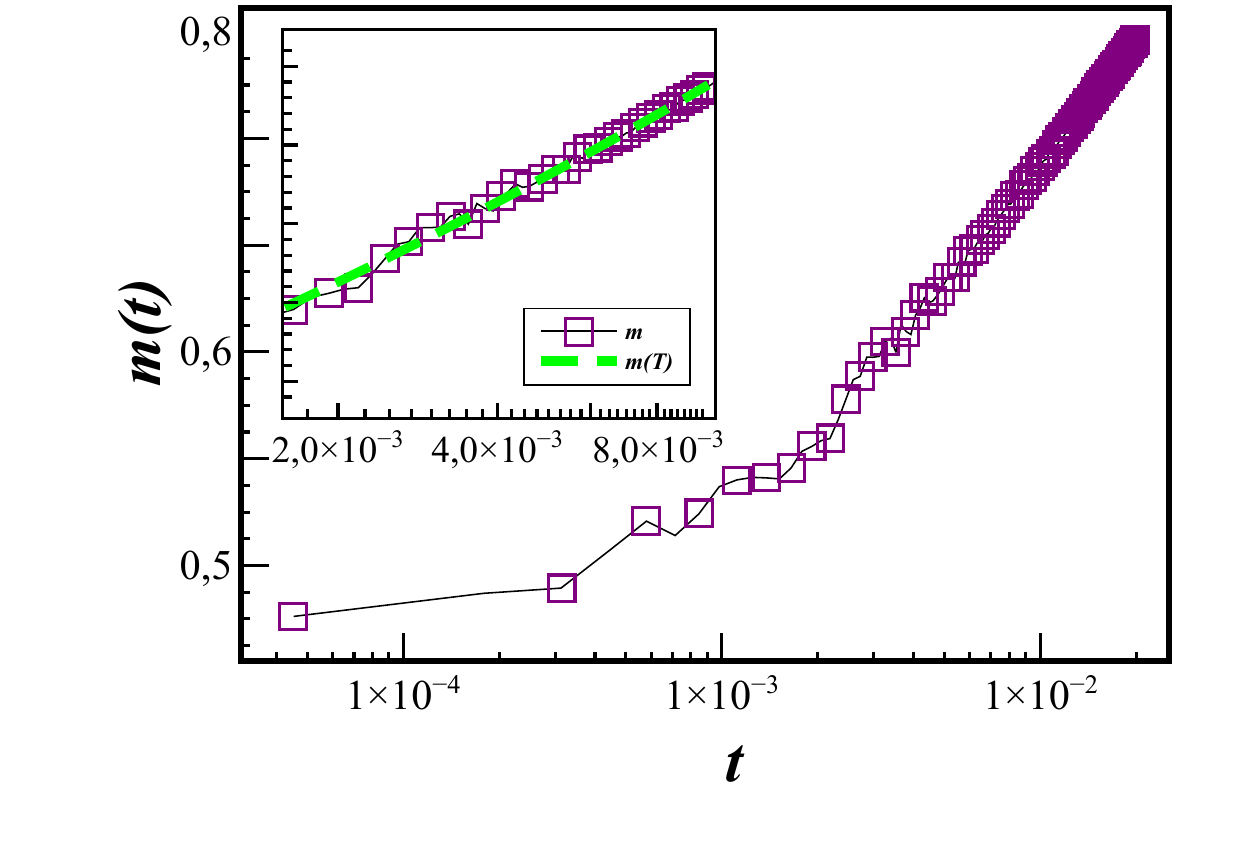}
		\caption{}
		\label{mag_fit_1}
	\end{subfigure}

	\caption{
		Sequential steps used to determine the transition temperature and the effective exponent $\beta$ for the $\sigma=0$ system.
		(a) Order-parameter curve $m(T)$ illustrating the choice of fitting windows. The lower bound $T_{\min}$ is kept fixed, while the upper bound $T_{\max}$ is varied to analyze the sensitivity of the fit. (b) Estimated critical temperature $T_c(T_{\max})$ as a function of the upper fitting bound. The highlighted region indicates the stability interval used to determine $T_c$.
		(c) Log--log plot of the order parameter $m(t)$, with $t\equiv (T_0-T)/T_0$, near criticality. The power-law fit gives the effective estimate $\beta=0.1251(9)$.}
	\label{fig:tc_beta_combined}
\end{figure}

Figure \ref{fig:tc_beta_combined} illustrates the procedure used to determine the transition temperature and the effective exponent estimates. We first extract the critical temperature $T_c$ from the order-parameter curve $m(T)$ for each value of the disorder strength $\sigma$. Data are fitted using Eq. (\ref{box1}), with $a$, $\beta$, and $T_c$ as fitting parameters. Since the fluctuations increase near the transition and the simulated systems are finite, the resulting estimates depend on the fitting interval. This point is particularly important in the present problem because logarithmic corrections are expected in the two-dimensional random-bond Ising model \cite{Reis96,Reis97,Fytas10}.

Figure \ref{mag_tc_tmax} shows the magnetization curve for $\sigma=0$ and illustrates the use of different adjustment windows. We keep $T_{\min}$ fixed and vary $T_{\max}$, obtaining an estimate of $T_c$ for each interval $[T_{\min},T_{\max}]$. The dependence of the fit critical temperature on $T_{\max}$ is shown in Fig. \ref{tc_tmax}. We then identify a stability region, highlighted in the figure, from which the final estimate of $T_c$ is obtained. The critical temperature determined from this magnetization analysis is denoted by $T_c^{\mathrm{mag}}$.

Using $T_c^{\mathrm{mag}}$, we estimate the effective exponents $\beta$ and $\nu$ from the leading powers-law forms $	m(T)\sim a|T-T_c|^{\beta}$ and $\rho(T)\sim b|T-T_c|^{-\nu}$ for different values of the disorder $\sigma$. 
To reduce finite-size effects, we repeated the analysis for the lattices with $L=1024m$, with $m=1,2,3,4,$ and $5$, and used a simple finite-size scaling to improve the exponent estimates reported in Table \ref{Table1}. Even with this procedure, the values of the individual exponents remain sensitive to the fitting method, as expected for a marginally disordered system \cite{Reis97,Fytas10}.

\begin{figure}[H]
	\centering

	% Painel (a)
	\begin{subfigure}[h!]{0.93 \linewidth}
		\centering
		\includegraphics[width=\linewidth]{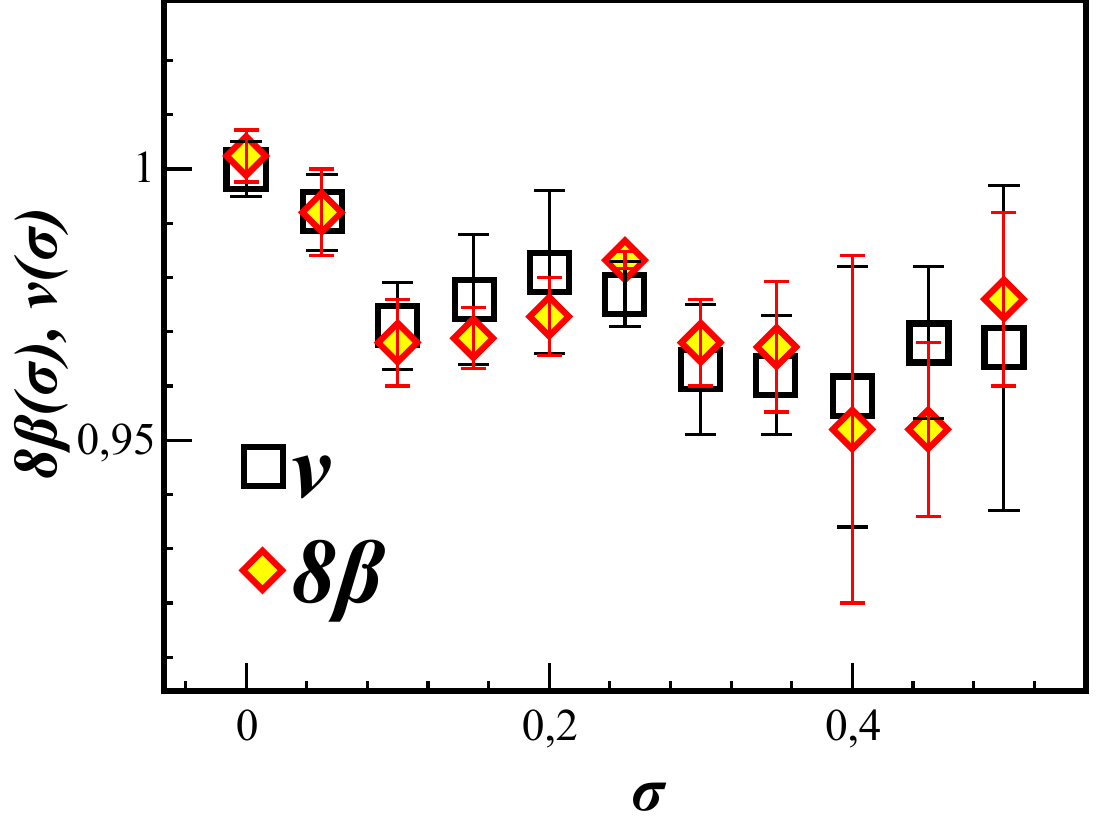}
		\caption{}
		\label{beta_nu}
	\end{subfigure}
	% Painel (b)
	\begin{subfigure}[h!]{0.93 \linewidth}
		\centering
		\includegraphics[width=\linewidth]{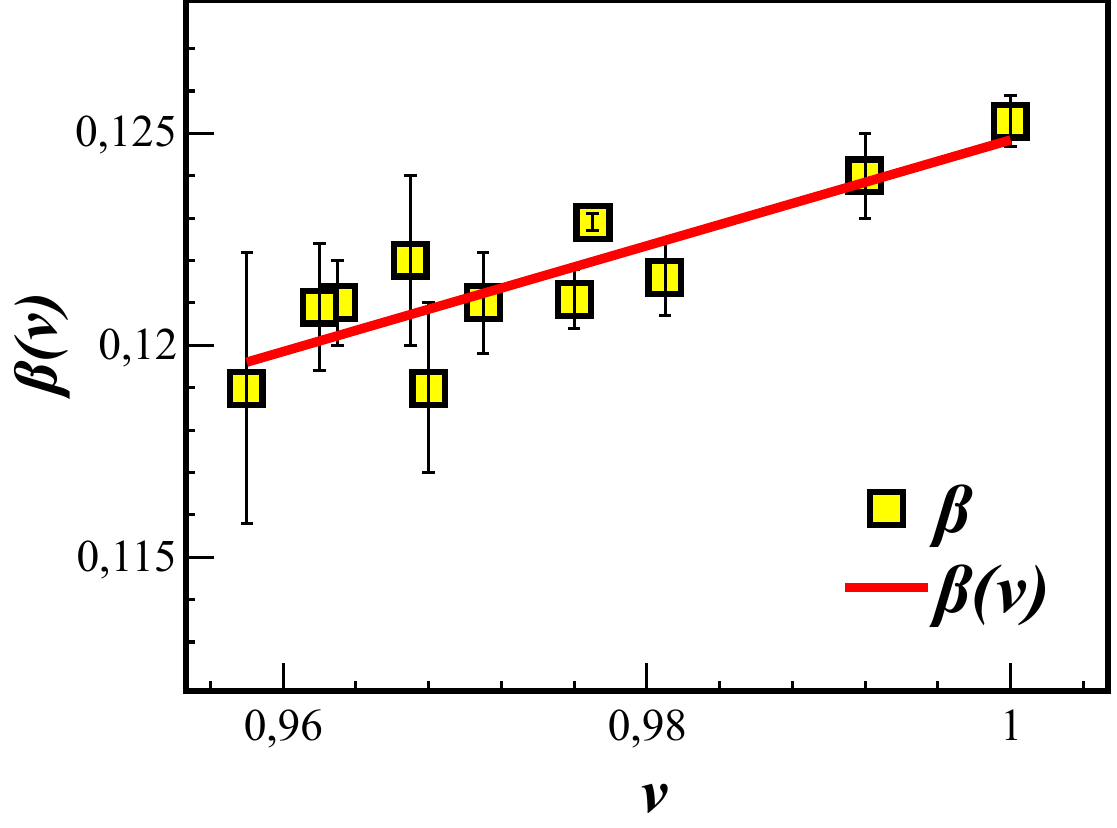}
		\caption{}
		\label{beta_nu_relation}
	\end{subfigure}
	\caption{ Evolution of the effective exponent estimates $\beta$ and $\nu$ as functions of disorder.
		(a) $\nu$ and $8\beta$ as functions of the standard deviation $\sigma$. The values of $\beta$, including their uncertainties, are multiplied by $8$ to allow a direct comparison with $\nu$.
		(b) Effective exponent $\beta$ as a function of $\nu$. The red line shows the linear relation $\beta=a\nu$, with $a=0.1248(3)$, consistent with the pure Ising ratio $\beta/\nu=1/8$.}
	\label{fig:beta_nu_analysis}
\end{figure}

Figure \ref{fig:beta_nu_analysis} shows the behavior of $\beta$ and $\nu$ as the disorder strength is varied. In Fig. \ref{beta_nu}, we compare $\nu$ with $8\beta$, so that both quantities can be displayed on the same scale. The approximate agreement between $\nu$ and $8\beta$ indicates that the ratio $\beta/\nu$ remains close to the pure Ising value $1/8$. Figure \ref{beta_nu_relation} shows $\beta$ as a function of $\nu$. The data are well described by a linear relation $\beta=a\nu$, with $a=0.1248(3)$, again consistent with $\beta/\nu=1/8$.

\begin{figure}[h!]
	\centering
	\includegraphics[width=0.94\linewidth]{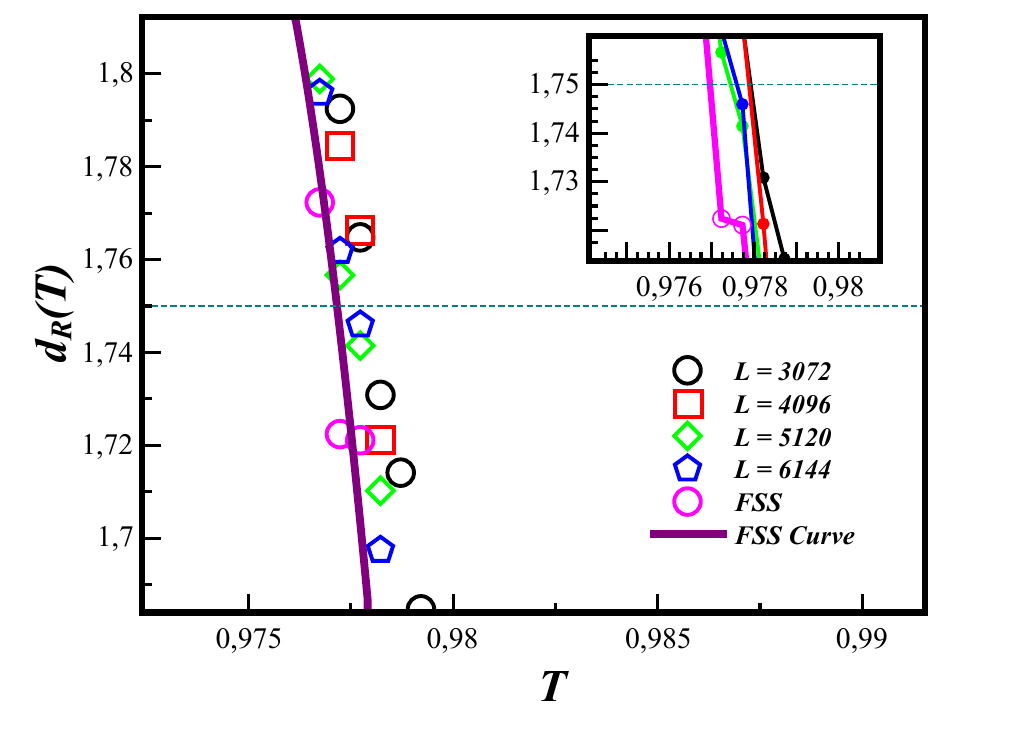}
	\caption{Normalized Riesz fractal dimension $d_R$ as a function of the  temperature $T/T_O$ for some values of $L$, and a standard deviation $\sigma = 0.25$.  We do a FSS for each value of the temperature to get a smoother $d_R(T)$, presented by the magenta curve. The dashed line indicates the exact Ising value $d_R=7/4$.}
	\label{fig:dr2}
\end{figure}

Figure \ref{fig:dr2} shows $d_R$ as function of $T$ for some values of $L$, and $\sigma=0.25$.  Here, we perform a finite size scaling for each temperature to smooth the curve so that, near the transition, it is possible to adjust the data using
\begin{equation}
	\label{drT}
	d_R(T)=d_R^* + C_1(T-T_c^{d_R})+C_2(T-T_c^{d_R})^{2},
\end{equation}
 where $d_R^*$ and $T_c^{d_R}$ are respectively the Riesz fractal dimension  and transition temperature estimated from these points. From this  we get $d_R^*= 1.759(19)$ and $T_c^{d_R}= 0.97704(23)$, which allows us to trace the continuous curve.  Similar data for each sigma enable us to obtain these parameters as a function of disorder. The comparison between $T_c^{\mathrm{mag}}$ and $T_c^{d_R}$ provides a useful estimate of the systematic uncertainty associated with the two procedures  and shows that the measurement of $d_R$ agrees with the theoretical value $d_R=7/4$.

\begin{table}[h!]
	\centering
	\begin{tabular}{cccc}
		\hline\hline
		$\sigma$ & $d_f$ & $d_R$ & $d_R^{*}$ \\
		\hline
		0    & 1.8747(6)  & 1.7494(13) & 1.75(5)   \\
		0.05 & 1.8750(13) & 1.7500(27) & 1.751(6)  \\
		0.10 & 1.8754(15) & 1.7508(29) & 1.759(12) \\
		0.15 & 1.8759(17) & 1.7518(34) & 1.757(26) \\
		0.20 & 1.8760(21) & 1.7521(42) & 1.76(8)   \\
		0.25 & 1.8742(8)  & 1.7484(16) & 1.759(19) \\
		0.30 & 1.8744(19) & 1.7487(38) & 1.77(4)   \\
		0.35 & 1.8743(21) & 1.7486(42) & 1.75(1)   \\
		0.40 & 1.8758(52) & 1.752(10)  & 1.76(3)   \\
		0.45 & 1.8771(27) & 1.7541(55) & 1.753(14) \\
		0.50 & 1.874(4)   & 1.748(9)   & 1.750(6)  \\
		\hline\hline
	\end{tabular}
	\caption{
		Fractal dimensions $d_f$, $d_R$, and $d_R^{*}$ as functions of the disorder strength $\sigma$. The quantities $d_f$ and $d_R$ are obtained from the exponent ratio $\beta/\nu$ using Eqs.~(\ref{df}) and (\ref{dR}), respectively. The quantity $d_R^{*}$ corresponds to the fractal dimension obtained from the direct fractal analysis of $d_R(T)$ from Eq.~(\ref{drT}). All estimates fluctuate around, and remain compatible within uncertainties with, the exact pure Ising values $d_R=7/4$ and $d_f=15/8$, indicating that the geometrical exponents are more stable than the individual effective estimates of $\beta$ and $\nu$.
	}
	\label{Table2}
\end{table}

The fractal dimensions are summarized in Table \ref{Table2}. In this table we compare two different routes to estimate the geometrical exponents. The first one is based on the exponent ratio $\beta/\nu$: using Eq. (\ref{df}), we obtain $d_f=d_f(\beta/\nu)$, and then Eq. (\ref{dR}) gives the corresponding value $d_R=d_R(\beta/\nu)$. Since $\beta/\nu$ remains close to the pure Ising value $1/8$, these two quantities are expected to remain close to $d_f=15/8$ and $d_R=7/4$, respectively. The second route is independent of the thermodynamic exponent estimates. In this case, $d_R^{*}$ is obtained directly from the geometrical analysis of the curves $d_R(T)$, using Eq. (\ref{drT}), see Figure (\ref{fig:dr2}). As shown in Table \ref{Table2}, the directly measured values $d_R^{*}$ remain close to the pure Ising value $7/4$ over the whole range of disorder investigated, with no systematic drift as $\sigma$ increases. This result contrasts with the stronger disorder-dependent fluctuations observed in the individual effective estimates of $\beta$ and $\nu$ in Table \ref{Table1}. Thus, while pure power-law fits of thermodynamic quantities are sensitive to logarithmic corrections, finite-size effects, and fitting-window choices, the fractal dimension $d_R$ provides a more stable geometrical signature of the underlying universality class. In particular, the strong-universality scenario is obtained here from a comparatively simple geometrical analysis, despite the limited precision of the individual exponent estimates, whereas previous numerical studies have relied on high-precision transfer-matrix, Wang--Landau, or detailed finite-size scaling approaches to resolve the same universality issue \cite{Reis96,Reis97,Fytas08,Fytas10}.

\section{Geometrical observables and strong universality}

%Ismael: Pra ser sincero, acho que essa seção virou uma pré conclusão que só introduz redundância. Acho melhor puxar parte para os resultados e jogar o resto para a conclusão (se já não houver discussões análogas por lá)

The above results should be interpreted in the context of marginal disorder in the two-dimensional random-bond Ising model. Quenched fluctuations in the couplings preserve the global $Z_2$ symmetry of the Ising Hamiltonian and the short-range character of the interactions. According to the Harris criterion, bond disorder therefore acts as a marginal perturbation of the pure two-dimensional Ising fixed point, for which logarithmic corrections are expected \cite{Harris74,Reis96,Reis97}. Pure power-law fits performed for finite systems and finite fitting windows may produce disorder-dependent effective exponents even when the asymptotic critical behavior remains unchanged \cite{Reis96,Reis97,Fytas10}.

A more robust signature is provided by exponent ratios and the associated geometrical quantities. Although the individual estimates of $\beta$ and $\nu$ variates with the disorder strength, their ratio $\beta/\nu$ remains close to the pure Ising value $1/8$. Through the scaling relation $d_f=d-\beta/\nu$, Eq. (\ref{df}), this implies a nearly disorder-independent fractal dimension $d_f$. In turn, Eqs. (\ref{dR}) and (\ref{eta}) connect this geometrical quantity to the Riesz fractal dimension $d_R$ and to the anomalous exponent $\eta$.

Importantly, the stability of $d_R$ is not only inferred from the ratio $\beta/\nu$. Direct box count analysis provides an independent estimate of $d_R$, which remains consistent with the pure Ising value $d_R=7/4$ throughout the range of the investigated disorder and does not show a systematic deviation from $\sigma$. Its stability is comparable to that obtained indirectly from $\beta/\nu$, despite the larger uncertainties inherent in the box-counting estimate, which is usually more sensitive to finite-size effects and to the accessible range of length scales. Therefore, $d_R$ provides a comparatively direct and robust geometrical signature of strong universality without requiring a advanced methods to determine of the individual exponents.

The agreement between the thermodynamic and direct geometrical routes also provides further support for the fractal mean-field description and for the relations connecting $d_f$, $d_R$, and $\eta$ \cite{Lima24,Lima25,Carrasco26}.

\section{Concluding remarks and future perspectives}

{\bf Concluding remarks }
In this work, we investigate the critical behavior of the two-dimensional random-bond Ising model with Gaussian-distributed positive mean couplings. We analyze the dependence of the order parameter, correlation length, effective critical exponents, and fractal dimensions of the strength  disorder $\sigma$.

Direct estimates of  individual exponents $\beta$ and $\nu$ show noticeable variations as $\sigma$ increases. This behavior is consistent with the expected difficulty of extracting asymptotic exponents in the presence of marginal disorder, where logarithmic corrections and finite-size effects may lead to effective disorder-dependent estimates when pure power-law fits are used.

In contrast, the ratio $\beta/\nu$ remains close to the pure two-dimensional Ising value $1/8$. This stability is reflected in the fractal dimensions obtained from the scaling relations involving $d_f$ and $d_R$. More importantly, the direct estimates of $d_R$ obtained from the geometric analysis of the curves $d_R(T)$ also remain close to the pure Ising value $7/4$, with fluctuations comparable to those obtained from the exponent-ratio route. Thus, the robustness of $d_R$ is not only a consequence of rewriting $\beta/\nu$ in geometrical form, but is also observed directly in the geometry of the critical configurations.

These results support the strong-universality scenario with logarithmic corrections for the two-dimensional random-bond Ising model. Importantly, the direct box-counting estimate of $d_R$ shows no systematic drift with disorder and remains compatible with $7/4$. Thus,  the Riesz fractal dimension provides a comparatively simple and  stable geometrical signature of the underlying universality class when conventional exponent estimates are affected by scaling corrections.

Previous works showed that the fractional-geometrical formulation based on $d_R$ provides a consistent description of Ising criticality in integer and non-integer dimensions~\cite{Lima25,Carrasco26}, and for other models~\cite{Souza26} as well. The present results extend this picture by showing that the same geometrical relations involving $d_f$, $d_R$, and $\eta$ remain robust in the presence of quenched bond disorder. In this sense, the fractal-geometrical description remains compatible with the strong-universality scenario and provides a useful complementary route for identifying universality in marginally disordered systems.

{\bf Future perspectives beyond equilibrium -} Considering that out-of-equilibrium growth phenomena exhibit analogous relationships, see \cite{Anjos21,GomesFilho21b,GomesFilho24}, we infer that a general underlying relationship may exist, provided each case's fractal dimension is well-defined. Therefore, we propose that this hypothesis be rigorously tested across a variety of systems, such as those with a finite geometry\cite{GomesFolho16,Kalosakas22,GomesFilho22,salman24}, dynamic phase transitions~\cite{Ziff86,Fernandes18,Santos24} and synchronized phase transitions~\cite{Kuramoto84,Pinto16,Pinto17,Gutierrez24,GUTIERREZ25}. In the latter, once the synchronized state is reached, the order parameter becomes constant, exhibiting a behavior similar to magnetization\cite{Kuramoto84}.  Another important investigation concerns non-Hermitian dynamics~\cite{Kawabata19,Koch22}, where $J_{i,k} \neq J_{k,i}$	which enriches the topological phases beyond the existing Hermitian structure.

{\bf Acknowledgments -} This work was supported by 
the Conselho Nacional de Desenvolvimento Cient\'{i}fico e Tecnol\'{o}gico (CNPq), Grant No.  303119/2022-5, and Funda\c{c}\~ao de Apoio a Pesquisa do Distrito Federal (FAPDF), Grant No.\ 00193-00001817/2023-43. As well we thanks Prof. Dr. Marcio Sampaio for providing the computacional resources used in this work.
%\end{acknowledgments}

\newpage

\bibliography{ReferencesV2}

\end{document}